 \PassOptionsToPackage{table,xcdraw}{xcolor}
\documentclass[acmlarge]{acmart}

\AtBeginDocument{%
  \providecommand\BibTeX{{%
    \normalfont B\kern-0.5em{\scshape i\kern-0.25em b}\kern-0.8em\TeX}}}

\usepackage{graphicx}
\usepackage{array}
\usepackage{bbold}
\usepackage{caption}
\usepackage{subcaption}
\usepackage{multirow}

\usepackage{array}
\usepackage[table]{xcolor}
\renewcommand\footnotetextcopyrightpermission[1]{} 
\begin{document}

\title{Host-Based Network Intrusion Detection via Feature Flattening and Two-stage Collaborative Classifier}

\author{Zhiyan Chen}
\email{zchen241@uottawa.ca}
\author{Murat Simsek}
\email{murat.simsek@uottawa.ca}
\author{Burak Kantarci}
\email{burak.kantarci@uottawa.ca}
\affiliation{%
  \institution{University of Ottawa}
  \streetaddress{800 King Edward Avenue}
  \city{Ottawa}
  \state{Ontario}
  \country{Canada}
  \postcode{K1N 6N5}
}

\author{Mehran Bagheri}
\email{mbagheri@ciena.com}
\author{Petar Djukic}
\email{pdjukic@ciena.com}
\affiliation{%
  \institution{(were with) Ciena}
  \streetaddress{383 Terry Fox}
  \city{Ottawa}
  \country{Canada}
  \postcode{K2K 0L1}
}

\begin{abstract}
Network Intrusion Detection Systems (NIDS) have been extensively investigated by monitoring real network traffic and analyzing suspicious activities. However, there are limitations in detecting specific types of attacks with NIDS, such as Advanced Persistent Threats (APT). Additionally, NIDS is restricted in observing complete traffic information due to encrypted traffic or a lack of authority.
To address these limitations, a Host-based Intrusion Detection system (HIDS) evaluates resources in the host, including logs, files, and folders, to identify APT attacks that routinely inject malicious files into victimized nodes. In this study, a hybrid network intrusion detection system that combines NIDS and HIDS is proposed to improve intrusion detection performance. The feature flattening technique is applied to flatten two-dimensional host-based features into one-dimensional vectors, which can be directly used by traditional Machine Learning (ML) models.
A two-stage collaborative classifier is introduced that deploys two levels of ML algorithms to identify network intrusions. In the first stage, a binary classifier is used to detect benign samples. All detected attack types undergo a multi-class classifier to reduce the complexity of the original problem and improve the overall detection performance.
The proposed method is shown to generalize across two well-known datasets, CICIDS 2018 and NDSec-1. The performance of XGBoost, which represents conventional ML, is evaluated. Combining host and network features enhances attack detection performance (macro average F1 score) by $8.1\%$ under the CICIDS 2018 dataset and $3.7\%$ under the NDSec-1 dataset.
Meanwhile, the two-stage collaborative classifier improves detection performance for most single classes, especially for DoS-LOIC-UDP and DoS-SlowHTTPTest, with improvements of $30.7\%$ and $84.3\%$, respectively, when compared with the traditional ML XGBoost.
\end{abstract}



\received{6 May 2023}

\maketitle

\section{Introduction}
The importance of security solutions for networked systems has increased with the advances in information and communication technologies \cite{chen2022machine,zhang2022environmental}. To protect network systems, Network Intrusion Detection Systems (NIDSs) have been widely investigated and implemented \cite{yang2022systematic,de2023distributed}. ML-based NIDSs have been proven to detect prevalent and zero-day attacks, which have also been studied in the context of defensive and proactive/adversarial ML \cite{Liu.comst}. For instance, an XGBoost-DNN integrated NIDS \cite{devan2020efficient} has demonstrated decent performance.

NIDS identifies suspicious activities by analyzing data from a single packet in captured global network traffic \cite{lata2022intrusion,otoum.2021}. However, monitoring entire network traffic can be challenging in scenarios such as encrypted traffic or without authentication monitoring. NIDS cannot inspect encrypted network traffic, limiting its ability to detect only external attacks \cite{zipperle2022provenance}. Moreover, NIDS requires improved solutions to identify specific malicious activities, particularly APT, which often utilize malicious files attached to various applications \cite{moon2016host}.
Host-based intrusion detection evaluates resources in a host, including logs, files, and folders, to detect attacks on hosts such as servers. HIDS provides a fine-grained solution to detect anomalous patterns internally, making it a valuable tool in detecting APT \cite{mvula2023evaluating}. HIDS also has the advantage of detecting anomalies without analyzing or monitoring network traffic. As an example, \cite{ribeiro2020autonomous} presents a HIDS use case for securing Android mobile equipment. 
HIDSs are software components installed on observed systems, and they scan the entire system to prevent intrusions. HIDSs offer rich context, enabling excellent knowledge for data processing and analysis \cite{martins2022host}. Although HIDSs increase costs due to their connectivity to the server, setup of distributed clients, and collection and management of massive and sensitive data from host devices, they have recently gained attention from researchers. Industrial companies often implement both NIDS and HIDS to achieve promising detection performance and secure their systems \cite{rani2019review}.

\begin{figure}[ht]
    \centering
    \includegraphics[width = 0.7\textwidth, trim=0cm 0cm 0cm 0cm,clip]{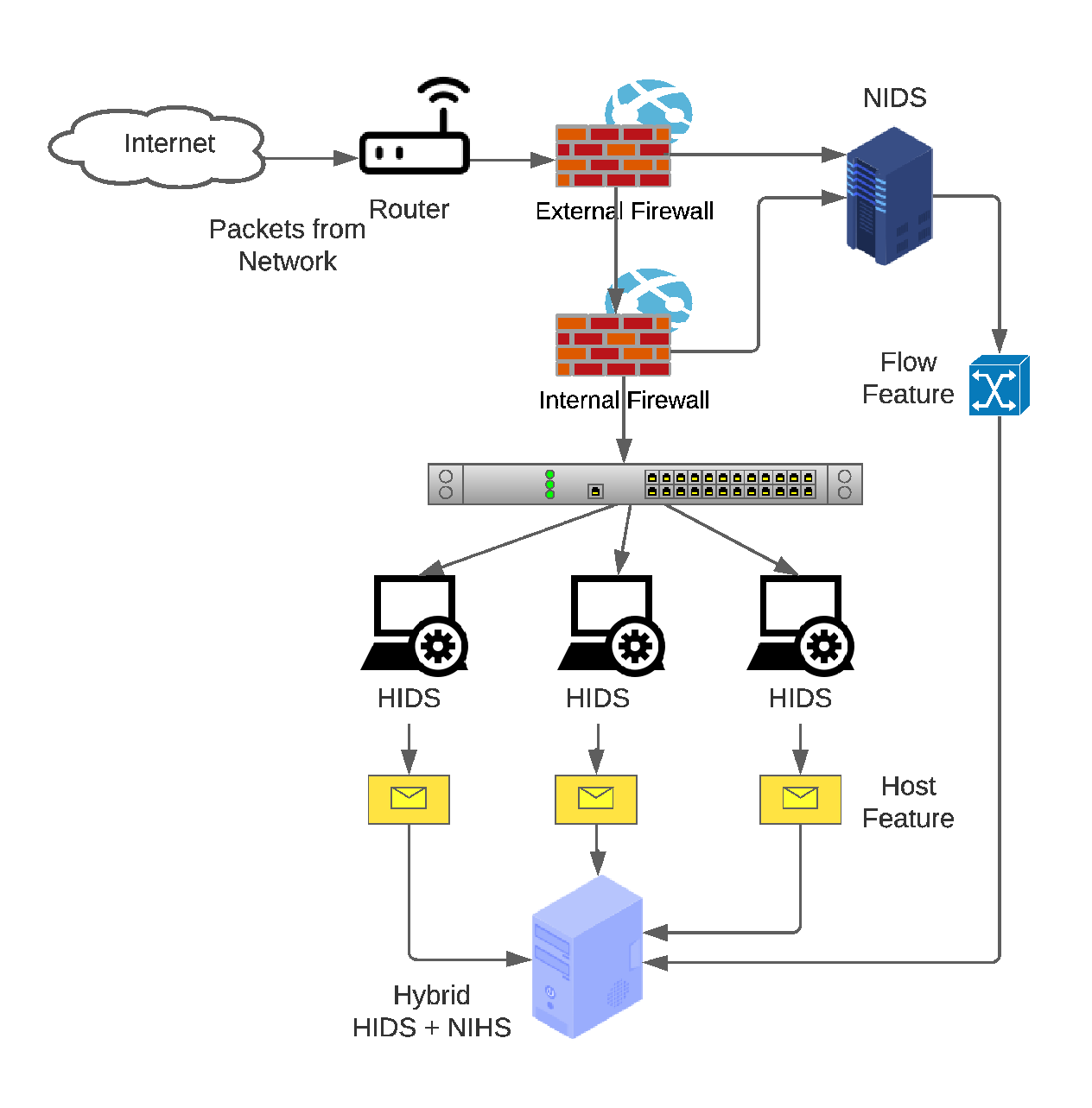}
    \caption{An example of hybrid HIDS and NIDS system }
    \label{fig:NIDS_HIDS_St}
\end{figure} 

In this work, we investigate intrusion detection system combining NIDS and HIDS via utilizing network and host features together. Various public datasets, such as NSL KDD, KDD 99, Bot-IoT, and MQTTTest, are utilized in network intrusion detection research \cite{chen2022machine}. However, there are only a limited number of public datasets that include host information. The CICIDS 2018 dataset and NDSec-1 dataset, which contain host-based information such as logs for events and messages, have been used in intrusion detection schemes \cite{stiawan2020cicids,nashat2021multifractal}. Therefore, we selected these two datasets to demonstrate the method integrating network and host features and evaluate the proposed method.
Despite the potential benefits of host-based information in intrusion detection, very few studies have applied it \cite{besharati2019lr}. This paper aims to bridge the gap and boost detection performance via analyzing host-resources. Host-based content is stored as a string and transformed into a numeric array by Bert \cite{prottasha2022transfer, kowsher2022bangla}. Event data and message data stored in the host are transformed separately to obtain host-based features. Moreover, the transformed host features are applied data preprocessing techniques to reduce the size of host features and finally are flattened from two-dimensional into a vector. The following step is to combine network/flow features \footnote{We use network and flow feature to represent the same feature in this work that are extracted from network traffic data and different with host features.} and host features and send to detection model, including traditional ML and the two-stage collaborative classifier.


One of the motivations behind this work is the large number of samples in intrusion datasets. For example, the CICIDS 2018 dataset contains approximately 1 million samples. Typically, the majority of the dataset consists of benign traffic, and attack samples are a minority since only a limited number of attack points can be appropriately hidden (except for denial-of-service related attacks). As a result, the large number of benign samples increases the complexity of the intrusion detection system due to the increased training overhead of the ML models. To address this, this work leverages both network-based and host-based intrusion detection frameworks simultaneously to take advantage of both systems. A two-stage collaborative classifier is proposed to improve intrusion detection accuracy and minimize false alarm probability. 
Fig. \ref{fig:NIDS_HIDS_St} illustrates the proposed intrusion detection system, which bridges HIDS and NIDS to consider both real network traffic and records saved in host devices. To reduce detection system complexity while maintaining detection performance, this article introduces a two-stage collaborative classifier that comprises a binary classifier and a multi-classifier. The binary classifier discriminates between benign and attack instances, with all attacks sharing the same label. The multi-class classifier characterizes the attack instances after the binary classifier has filtered out the benign instances. The proposed framework initially eliminates benign traffic and then focuses on recognizing individual attack categories.

Numerical results demonstrate improved attack detection performance under HIDS and NIDS when compared to NIDS alone. Using the CICIDS 2018 dataset, the detection performance of all individual attack classes is improved. For instance, the F1 score of DDOS-LOIC-UDP is improved from $0.7609$ to $0.9942$, while the F1 score for DoS-SlowHTTPTest is improved from $0.5425$ to nearly $1$. Moreover, the overall performance (macro average F1 score) improves significantly, from $0.9246$ to $0.9993$, with an $8.1\%$ improvement. Similarly, using the NDSec-1 dataset, incorporating host features increases the overall detection performance up to $0.8913$ in terms of macro average F1 score, while the macro F1 score is $0.8595$ when only flow features are used.
The evaluation results of the two-stage collaborative classifier show that the overall performance improves up to $0.9994$ under the CICIDS 2018 dataset. Moreover, the two-stage collaborative classifier improves detection performance for most individual attack classes, especially for DoS-LOIC-UDP and DoS-SlowHTTPTest, with improvements of $30.7\%$ and $84.3\%$, respectively, when compared to the traditional ML XGBoost. 
The main contributions of this work can be listed as below:
\begin{itemize}
    \item To the best of our knowledge, this paper integrates network features and host features and applies them to ML algorithms for training and prediction in intrusion detection. The presented method has been shown to improve intrusion detection performance when compared with traditional network feature-based detection approaches.
    \item The data flattening technique is used to map multidimensional features into a vector that can be easily utilized by ML or deep learning networks for training and testing.
    \item A two-stage collaborative classifier is introduced for network and host features, which reduces computation complexity and improves performance compared to conventional ML algorithms.
    
\end{itemize}

The remainder of the article is structured as follows. Section \ref{sec:background} presents the state-of-the-art in network security, emphasizing NIDS and HIDS. Section \ref{sec:dataset_intro} introduces two public datasets, CICIDS 2018 and NDSec-1, and provides a sample distribution for each. In Section \ref{sec:flat_method}, a machine learning-based approach using hybrid and network-based features is presented for detecting network intrusions, demonstrating the benefits of using host features alongside flow features. Section \ref{sec:cas} introduces a two-stage machine learning framework, which filters out benign samples before characterizing the attack instances. Finally, the article concludes in Section \ref{sec:conclusion} by discussing future directions.

\section{Related Work}
\label{sec:background}
Network security encompasses more than just communication networks; it also encompasses the security of the entire cyberspace environment, including the information infrastructure, application systems, and data resources \cite{xu2022network}. To achieve comprehensive network system security, several approaches can be employed, such as firewalls, antivirus software, access control, and anti-malware \cite{8962096, vacca2013network, chen2022machine, liu2021risk}. For instance, configuration and path analysis can help safeguard critical infrastructure, such as a smart-grid network, and can be viewed as a firewall use case \cite{tyav2022comprehensive}.
One study \cite{al2022multiclass} proposes a neural network-based classification system that uses firewall logs to ensure system security. NIDS have also become widely used for intrusion detection, particularly when combined with machine learning algorithms, which have shown promising detection performance. Another study \cite{fouladi2022ddos} presents a neural network-based framework that identifies abnormal actions in the system to protect against attacks. In \cite{han2020unicorn}, the authors introduce UNICORN, which uses time-efficient figures to extract provenance images, providing in-depth information to identify APT attacks. Finally, \cite{mendoncca2022lightweight} presents a deep learning-based attack detection system for industrial internet of things, called the Sparse Evolutionary Training Approach. The proposed method demonstrates promising detection performance.
One study \cite{ravi2022recurrent} proposes a deep learning-integrated recurrent approach to detect network intrusions by extracting features from recurrent models. The approach also employs a feature selection technique to identify optimal features from the extracted and original ones, and an ensemble model for classification. Another study \cite{liu2020machine} presents a NIDS framework that can detect specific attacks by integrating supervised (such as XGBoost) and unsupervised (such as expectation-maximization) machine learning algorithms.
The study in \cite{santos2019clustering} presents an intrusion detection system to prevent routing attacks such as sinkhole and selective forwarding attacks in IoT networks. The studies in \cite{chen2021all} demonstrate ensemble learning-based NIDS that take advantage of various individual ML models to estimate wisely. It is worth to note that there have been studies that leverage federated learning-based intrusion detection \cite{friha20232df} however, we scope this work to centralized ML-based intrusion detection solutions.

\begin{table}[!htb]
\begin{center}
\caption{List of existing works for network intrusion detection}
\centering
\label{tab:relatedwork}
\scalebox{0.92}{
\begin{tabular}{|m{1.5em}|m{10em}|m{6em}|m{10em}|m{10em}|m{4em}|}
\hline
Ref.                             & Methodology                                                                                                                         & Dataset                                                                          & Benefit       &    Further Improvement                                                                                                 &        Features                                                                                                                    \\ \hline
\cite{al2022multiclass}         & Identify intrusions in firewall via SNN                             & IFW-2019                     & 98.50\% accuracy                                                    & Dataset uniformity                                            & Flow          \\ \hline
\cite{fouladi2022ddos}          & DDoS identification using DWT and auto-encoder                      & Private                      & High detection rate 100\%                                           & Testing in a real SDN network                                 & Flow          \\ \hline
\cite{han2020unicorn}           & APT attach detection via a detector UNICORN                         & DARPA, CADETS, etc.          & Detect real-time APT attack                                         & Graph and host behavior analysis                              & Graph feature \\ \hline
\cite{mendoncca2022lightweight} & SET integrated attack detection model in IIoT                       & CICIDS 2017, DS2OS           & 2.2 ms average test time                                            & Diversity of test equipment                                   & Flow          \\ \hline
\cite{ravi2022recurrent}        & Via a end-to-end approach to identify intrusions                    & SDN-IoT, KDD 99, etc.        & Generalization                                                      & Testing in a real-time system                                 & Flow          \\ \hline
\cite{liu2020machine}           & Specific intrusions are investigated via ML-based NIDS              & NSL-KDD                      & Multiple ML models                                                  & Extensive testing of unsupervised algorithms                  & Flow          \\ \hline
\cite{santos2019clustering}     & Detection for sinkhole and selective forwarding via THATACHI        & Simulate via Cooja           & 50\% less energy consumption                                        & Investigation of other intrusions                             & N/A           \\ \hline
\cite{chen2021all}              & Ensemble learning-based NIDS                                        & NSL-KDD                      & Wise prediction is made                                             & Integration of other ML models                                & Flow          \\ \hline
\cite{el2022contextualizing}    & Intrusion detection via HIDS using two datasets                     & CB-DS, LID-DS                & Reasonable run-time overhead                                        & Identifying Heartbleed and a few other real-time intrusions.  & Host          \\ \hline
\cite{harshitha2020novel}       & Observe logs in hosts and identify illegitimate logs via OSSEC tool & N/A                          & Identify malicious logs                                             & Further test result analysis                                  & Host          \\ \hline
\cite{ribeiro2020autonomous}    & Statistical and semi-supervised ML-based HIDS                       & Private                      & Run on the mobile equipment and entirely autonomous                 & Extending datasets with more samples                          & Host          \\ \hline
\cite{martinez2021host}         & HIDS to secure embedded industrial equipment                        & N/A                          & Deploy HIDS firstly in a PLC, that has a real-time operation system & Extension of features and capability  of the presented system & Host          \\ \hline
\cite{prasad2022hidsc2}         & Two IDS merged to protect cloud                                  & N/A                          & Identify intrusions and create alerts                               & Combine different ML models                                 & Host, flow    \\ \hline
\cite{liu_collaborative_2022}   & Introduce CIDS-Net for intrusion detection                & CICIDS 2018, SCVIC-CIDS-2021 & 99.89\% macro F1 score                                              & Improving performance for some classes                        & Host, flow    \\ \hline
Our   & Hybrid NIDS and HIDS; Two-stage collaborative classifier           & CICIDS 2018, NDSec-1 & Use public dataset; boost performance                                            & Apply deep learning model                        & Host, flow    \\ \hline
\end{tabular}}
\end{center}
\end{table}

Related work discusses the use of host-based intrusion detection (HIDS) to detect specific attacks that are challenging for a NIDS. Additionally, HIDS eliminates the need for monitoring the entire network traffic by leveraging host resource analysis, making it appealing to researchers \cite{bridges2019survey}. 
For example, the study in \cite{el2022contextualizing} proposes a host-based system to detect network intrusion via monitoring and analyzing containerized environments (e.g., system calls), that is helpful to make accurate prediction for attacks. 
In \cite{harshitha2020novel}, the authors investigate observing information from system logs, which are host-based features, and show that such systems could classify suspicious activities with high accuracy. Furthermore, the authors in \cite{ribeiro2020autonomous} utilize host-based information to develop a detection system by organizing statistical data and ML models. However, there are few studies on hybrid network-based and host-based intrusion methods, such as the study in \cite{liu_collaborative_2022}. 
The study \cite{martinez2021host} demonstrates the use of HIDS in Industrial Automation Systems to identify intrusions targeting specific entities of embedded industrial equipment by analyzing information on host devices. 
In \cite{vinoth2022application}, a HIDS system integrated with Convolutional Neural Networks (CNN) is described to ensure security in IoT, with general characteristics that make it compatible with all IoT products. 
Another study \cite{prasad2022hidsc2} introduces HIDS to enhance cloud system security, integrating several machine learning algorithms (such as KNN, Logistic Regression, and Naive Bayes). The proposed HIDS detection system is efficient in handling large volume data in the cloud, in comparison to traditional NIDS systems. Table \ref{tab:relatedwork} presents current works to identify network intrusion. A thorough qualitative comparison in the table unveils the need for demonstrating a hybrid HIDS and NIDS.

While NIDS and HIDS are both effective in detecting various types of network intrusions, most current research tends to rely solely on one of them. Because there are limited ML-based frameworks that integrate both HIDS and NIDS to identify attack samples, a hybrid framework that leverages the strengths of both approaches would be valuable for improving detection performance. Additionally, previous studies have shown that machine learning algorithms are effective in building network intrusion detection systems. With this in mind, the goal of this work is to propose a machine learning-based hybrid approach that combines HIDS and NIDS to identify intrusions.

\begin{figure}[ht]
    \centering
    \includegraphics[width = 0.5\textwidth, trim=0.2cm 0.2cm 0.3cm 0cm,clip]{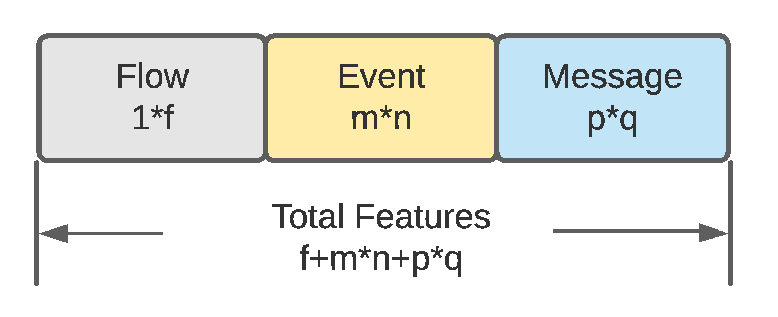}
    \caption{Flow features and host features after transforming. Initially presented in \cite{liu_collaborative_2022}. A sample consists of f flow, m*n event, and p*q message features after flattening.}
    \label{fig:flow_host_features}
\end{figure}

\section{Dataset introduction}
\label{sec:dataset_intro}
This work utilizes two public datasets that contain network and host information. Host-related data is text-based, extracted from host resources such as logs, files, and folders. The Transformer technique (Bert) is applied to convert text information into numerical data. Fig. \ref{fig:flow_host_features} illustrates the structure of network-based and host-based features in the CICIDS 2018 and NDSec-1 datasets. The message text and event text information are transformed into two-dimensional numerical matrices separately by Bert. Bert converts event text into an $m*n$ matrix and message text into a $p*q$ matrix. Similarly, the flow feature/network feature dimension is represented by a $1*f$ matrix, where $f$ is determined by the number of network features in the dataset. This transformation enables machine learning algorithms to utilize flow features and host features to train themselves, which benefits from both network intrusion detection systems and host-based intrusion detection systems \cite{liu_collaborative_2022}.

\begin{table}[!hbt]
\centering
\caption{CICIDS 2018 dataset sample distribution}
\label{tab:cicids_2017}
\begin{tabular}{|c|c|c|c|}
\hline
Class             & Training        & Test            & \textbf{Total}  \\ \hline
 Benign           & 308375          & 152172          & \textbf{460547} \\ \hline
 Bot              & 60767           & 29693           & \textbf{90460}  \\ \hline
 DDOS-HOIC        & 137147          & 67449           & \textbf{204596} \\ \hline
 DDOS-LOIC-HTTP   & 39019           & 19166           & \textbf{58185}  \\ \hline
 DDOS-LOIC-UDP    & 760             & 342             & \textbf{1102}   \\ \hline
DoS-GoldenEye    & 2271            & 1163            & \textbf{3434}   \\ \hline
 DoS-Hulk         & 13388           & 6553            & \textbf{19941}  \\ \hline
 DoS-SlowHTTPTest & 10579           & 5351            & \textbf{15930}  \\ \hline
 DoS-Slowloris    & 1394            & 702             & \textbf{2096}   \\ \hline
 FTP-BruteForce   & 32222           & 15918           & \textbf{48140}  \\ \hline
 SSH-Bruteforce   & 11143           & 5418            & \textbf{16561}  \\ \hline
\end{tabular}
\end{table}

\subsection{CICIDS 2018 dataset}
The CICIDS 2018 dataset is a public dataset that contains seven types of attacks, including Brute-force, Heartbleed, Botnet, DoS, DDoS, Web attacks, and infiltration of the network from inside \footnote{The dataset is introduced in detail at: https://www.unb.ca/cic/datasets/ids-2018.html}. The attacking infrastructure consists of $50$ devices, and the victim system has $5$ branches, comprising $420$ devices and $30$ servers. The CICIDS 2018 dataset comprises captured network traffic and recorded system logs of each device, along with $80$ features obtained from the captured traffic \cite{kanimozhi2019artificial}. The seven attack scenarios are categorized into 14 types, including Bot, DDOS-HOIC, DDOS-LOIC-HTTP, DDOS-LOIC-UDP, DoS-GoldenEye, DoS-Hulk, DoS-SlowHTTPTest, DoS-Slowingloris, FTP-BruteForce, SSH-Bruteforce, Brute Force-Web, Brute Force-XSS, Infiltration, and SQL Injection.
Since Brute Force-Web, Brute Force-XSS, Infiltration, and SQL Injection consist of very few samples (less than $50$), it is challenging to train machine learning and deep learning algorithms effectively. Compared to the significant number of other attack scenarios, the limited samples can be considered noise in the training procedure and result in lower detection performance. Therefore, the four classes are eliminated from the CICIDS 2018 dataset. The sample distribution of the remaining attack scenarios in the CICIDS 2018 dataset is presented in Table \ref{tab:cicids_2017}.

The CICIDS 2018 dataset contains network traffic with $132$ features that represent numerical arrays, which can be used directly by machine learning algorithms. This means that $f=132$ in Fig. \ref{fig:flow_host_features}. Most current research utilizes network-based data to build NIDS when using the CICIDS 2018 dataset. However, the CICIDS 2018 dataset also contains logs in the host with event data, message data, and network traffic. The logged data in host equipment is saved as text representation to ensure readability for clients and within the organization. To integrate network- and host-based data, text data needs to be transformed into numerical arrays. In this paper, we adopt the strategy proposed in \cite{liu_collaborative_2022} and use Bert for word embedding to extract host features, a transformer architecture trained to obtain language representations. We derive event data and message data separately. Table \ref{tab:matrix_CICID_NDSec} demonstrates that the CICIDS 2018 dataset consists of $132$ network features, $224$ event features, and $76,800$ message features.

\subsection{NDSec-1 dataset}
\label{sec:ndsec_dt_introduction}
The NDSec-1 dataset was initially introduced in \cite{Beer2017} and was generated for research in attack composition in network security schemes. The NDSec-1 dataset is a public dataset used in network intrusion detection \cite{nashat2021multifractal}\cite{liang2021co}. The dataset consists of 8 types of attacks, including Botnet, Bruteforce, DoS, Exploit, Malware, Misc, Probe, Webattack, Spoofing, and benign samples. However, there is only one sample for the Spoofing class, so this class is excluded from the evaluation section. Table \ref{tab:ndsec_DT_intro} presents the sample distribution across the training and test sets. Flow features are extracted from the NDSec-1 dataset and included with $63$ network features. Meanwhile, log records in hosts are provided, so host features are extracted using Bert for text-to-numeric array transforming. The CICIDS 2018 dataset's host features extraction procedures are adopted. Finally, the NDSec-1 dataset contains two parts of host features, including two-dimensional event and message features, as shown in Fig. \ref{fig:flow_host_features}. Table \ref{tab:matrix_CICID_NDSec} demonstrates that the NDSec-1 dataset consists of $63$ network features, $17,456$ event features, and $393,216$ message features.
\begin{table}[!hbt]
\centering
\caption{NDSec-1 dataset sample distribution}
\label{tab:ndsec_DT_intro}
\begin{tabular}{|c|c|c|c|c|}
\hline
Class      & Training & Testing & \textbf{Total} \\ \hline
BOTNET     & 52       & 40      & \textbf{92}    \\ \hline
 BRUTEFORCE & 2150     & 1006    & \textbf{3156}  \\ \hline
 DOS        & 12677    & 6389    & \textbf{19066} \\ \hline
 EXPLOIT    & 4        & 2       & \textbf{6}     \\ \hline
 MALWARE    & 23       & 12      & \textbf{35}    \\ \hline
 MISC       & 31       & 18      & \textbf{49}    \\ \hline
 NORMAL     & 5701     & 2284    & \textbf{7985}  \\ \hline
 PROBE      & 777      & 307     & \textbf{1084}  \\ \hline
 WEBATTACK  & 18       & 12      & \textbf{30}    \\ \hline
\end{tabular}
\end{table}

\begin{table}[!hbt]
\centering
\caption{Original dimension of CICIDS 2018 and NDSec-1 for flow, event, and message features}
\label{tab:matrix_CICID_NDSec}
\begin{tabular}{|c|c|c|c|}
\hline
            & Flow (1*f)   & Event (m*n)   & Message (p*q)   \\ \hline
CICIDS & 1*132  & 28*8  & 100*768 \\ \hline
NDSec-1     & 1*63  & 2182*8 & 512*768 \\ \hline
\end{tabular}
\end{table}

\begin{figure}[!hbt]
    \centering
    \includegraphics[width = 1.0\textwidth, trim=1.2cm 3.5cm 2cm 1.2cm,clip]{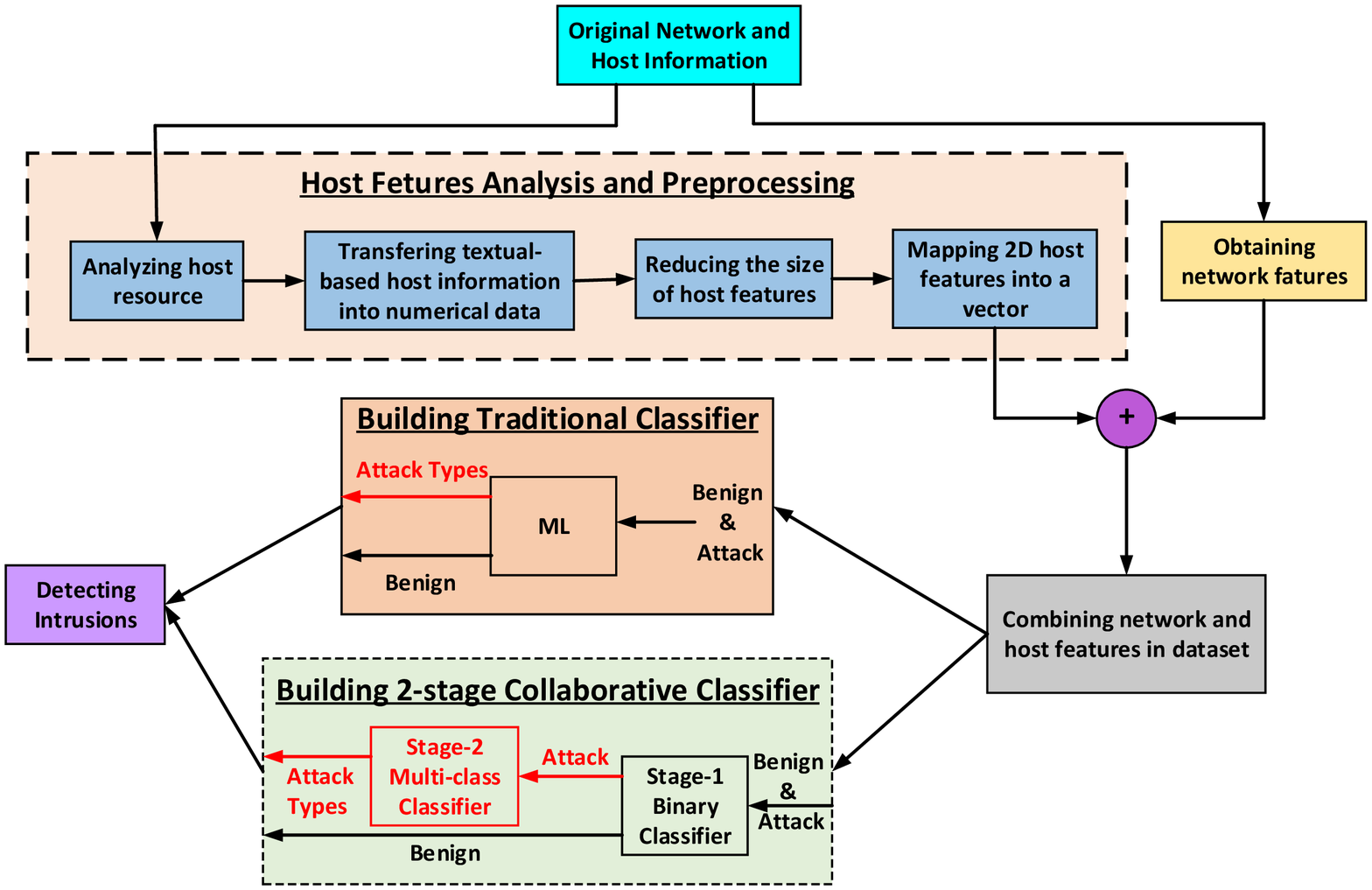}
    \caption{General framework for the proposed methodology. Algorithmic details of  two-stage Collaborative Classifier are given in Fig. \ref{fig:cascade_model_arch} }
    \label{fig:system_block_fig}
\end{figure}

\section{Hybrid network features and host features for intrusion detection}
\label{sec:flat_method}
This section explains the framework to combine network and host features for machine learning-integrated intrusion detection. Fig. \ref{fig:system_block_fig} demonstrates general framework of the hybrid network and host features that are utilized via ML-based detection approaches.

\subsection{Multiple dimension features combination for network and host features}
As introduced before, host-based features are extracted from event data and message information, which both have two dimensions as presented in Fig. \ref{fig:flow_host_features} (as presented in \cite{liu_collaborative_2022}). It is a problem to efficiently utilize the two-dimensional host features, like regular one-dimensional flow features. Feature flattening is a direct, easy, and low-cost technique to map a multi-dimension matrix into a vector. Feature flattening is used in CNN due to the lack of support for multidimensional data in the densely connected layers of a CNN \cite{abiwinanda2019brain}, \cite{albawi2017understanding}. Feature flattening is a feasible approach to adjusting multidimensional data to vectorized features. It is mainly used by ML algorithms.

A sample has $f$ flow features, which is represented as a vector  (\ref{equ:flow}).
\begin{equation}
\label{equ:flow}
\mathbf{A} = 
\begin{bmatrix}a_{11} & a_{12} &... & a_{1f}\end{bmatrix}
\end{equation}
Two-dimensional event features are denoted in (\ref{equ:event}).
\begin{equation}
\label{equ:event}
\mathbf{B} = 
\begin{bmatrix}
b_{11} & b_{12} &...&b_{1n}\\
b_{21} & b_{22} &... &b_{2j}\\
...&...&...&...\\
b_{n1} & b_{n2} &... &b_{mn}
\end{bmatrix}
\end{equation}
Finally, two-dimensional message features are represented in (\ref{equ:mes}).
\begin{equation}
\label{equ:mes}
\mathbf{C} = 
\begin{bmatrix}
c_{11} & c_{12} &... & c_{1q}\\
c_{21} & c_{22} &... &c_{2q}\\
... &...&...& ...\\
c_{m1} & c_{m2} &... &c_{pq}
\end{bmatrix}
\end{equation}
Flow features, also called network-based features, are the major features in NIDS, especially in ML-based detection systems \cite{devan2020efficient, albawi2017understanding, jiang2020network}. In order to utilize host-based features, two-dimensional feature data is flattened into a one-dimensional vector. As a result, after flattening, a vector $B^{'}$ of event-based features is obtained. 
\begin{equation}
\label{equ:event_flat}
\mathbf{B^{'}} = 
\begin{bmatrix}
b_{11} &... & b_{1n} &b_{21} &... &b_{2n} & ... &b_{mn} &... &b_{n8}\\
\end{bmatrix}
\end{equation}

Message-based host features are adjusted into a vector, denoted by $C^{'}$. 
\begin{equation}
\label{equ:event_flat}
\mathbf{C^{'}} = 
\begin{bmatrix}
c_{11} &...&c_{1q} & c_{21} &... c_{2q} &... &c_{pq}\\
\end{bmatrix}
\end{equation}
As a result, multi-dimensional host features are mapped to a vector, so it is easy to obtain hybrid features, denoted by $H_{1}$ combining flow features and event features, $H_{2}$ combining flow features and message features, and $H_{3}$ combining flow features, event features and message features, respectively.

\begin{equation}
\label{equ:flow_event_combine}
\mathbf{H_{1}} = 
\begin{bmatrix}
A & B^{'}\\
\end{bmatrix}
\end{equation}
\begin{equation}
\label{equ:flow_message_combine}
\mathbf{H_{2}} = 
\begin{bmatrix}
A &C^{'}\\
\end{bmatrix}
\end{equation}
\begin{equation}
\label{equ:flow_message_combine}
\mathbf{H_{3}} = 
\begin{bmatrix}
A & B^{'} &C^{'}\\
\end{bmatrix}
\end{equation}

Therefore, the number of hybrid features depends on the number of flow features, event features, and message features. It is straightforward to calculate the number of elements in vectors $B^{'}$ and $C^{'}$, which stand for event features ($t_{event}$) and message features ($t_{mes}$), respectively.

\begin{equation}
\label{equ:event_number}
t_{event} = m \times n
\end{equation}
\begin{equation}
\label{equ:event_number}
t_{mes} = p \times q
\end{equation}
Thus, $H_{3}$ contains the maximum number of features, represented in (\ref{equ:total}).
\begin{equation}
\label{equ:total}
t_{total} = f+ t_{event}+t_{mes}
\end{equation}

\subsection{Host features reduction}
In the CICIDS 2018 dataset, as introduced in Section \ref{sec:dataset_intro}, there are $132$ flow features, so $f$ is $132$ in $A$. Moreover, event features are extracted into a $288$ matrix, so $m$ equals $28$, and $n$ is $8$ in $B$. After flattening the event and message features, $B^{'}$ contains $t_{event}$, which equals $224$ elements, indicating that there are $224$ event features in vector $B^{'}$. Meanwhile, the message feature dimension is $100768$, represented by $C$, so $p$ is $100$, and $q$ is $768$ in $C$. After flattening the message feature, we obtain a vector $C^{'}$ consisting of $76800$ message features. The maximum number of features in the CICIDS 2018 dataset $t_{total}$ is $77,156$ features, according to Equation (\ref{equ:total}), including flow and host features.
Moreover, the CICIDS 2018 training dataset includes about $1$ million samples, as described in Table \ref{tab:cicids_2017}. Therefore, it is challenging to apply a total of $77,156$ features under our current test environment. Meanwhile, in the NDSec-1 dataset, as described in Section \ref{sec:ndsec_dt_introduction}, the number of flow features is $63$ ($f$). Event features are extracted into a $17,456$ matrix, and the message feature dimension is $393,216$. Thus, a sample in the NDSec-1 dataset consists of hybrid features up to $410,735$ when mixing flattened host features and network features, according to Equation (\ref{equ:total}). With an overwhelming number of features to be processed by an ML model, the NDSec-1 dataset faces the same dimensionality challenge as the CICIDS 2018 dataset.

With this in mind, this work reduces the message feature matrix and event feature matrix before flattening them into a vector. For CICIDS 2018 dataset, reduction of message features needs to be considered since message features constitute the majority (e.g., $76,800$) among all feature types. Shortening the message feature matrix from $100*768$ to a smaller matrix such as $10*10$ reduces message features directly. Furthermore, it is critical to reducing both the event feature matrix (i.e., $2182*8$) and host feature matrix (i.e., $512*768$) for the NDSec-1 dataset before flattening them into a vector array due to a large number of features for event and message features. 
A method must be designed to select a small matrix from the original large matrix and maintain intrusion detection performance. This work applies the random selection method to determine rows and columns for a matrix prior to feature flattening. The random selection method aims to ensure the selected message matrix is representative of the large one and less likely to be subject to bias \cite{carson1999random}. Specifically, message features in CICIDS 2018 are aimed to be reduced from $100*768$ to $10*15$; $10$ is randomly selected among the range $1$ to $100$, and $15$ is randomly selected from the range $1$ to $768$. It is worth noting that the average of several rounds of test results is measured to reduce the impact of fluctuations across samples and to take all features into account. We will demonstrate the determination of the appropriate dimension of a matrix considering computation performance and detection performance in Section \ref{sec:cicidsresults} and Section \ref{sec:results_NDSec}.

\begin{figure*}[!hbt]
    \centering
    \includegraphics[width = 1\textwidth, trim=0cm 0cm 0cm 0cm,clip]{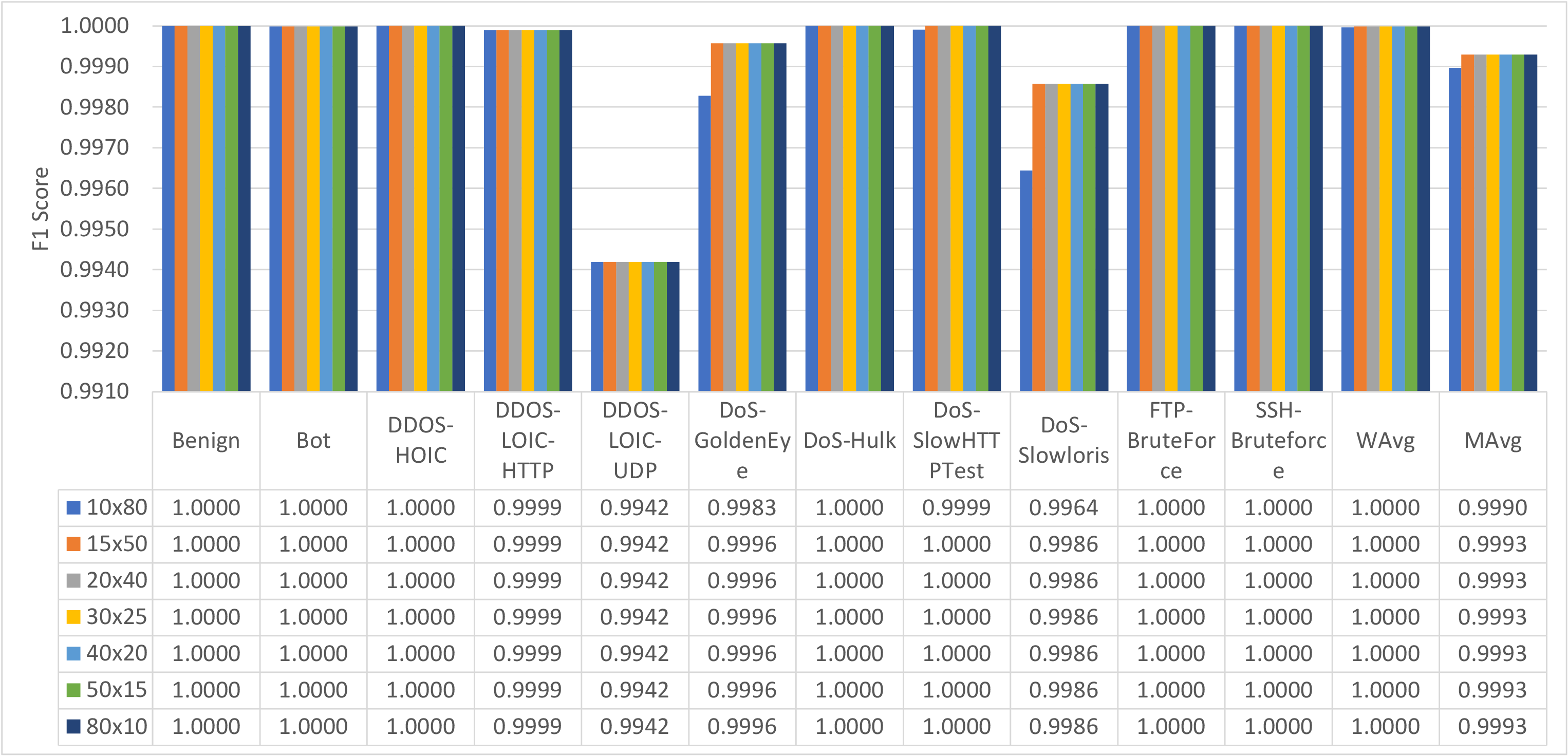}
    \caption{Performance comparison under various dimensions of message feature (host-based feature)}
    \label{fig:DT_diff_row_col}
\end{figure*}

\subsection{Performance evaluation under CICIDS 2018 dataset}
\label{sec:cicidsresults}
As stated before, the message features matrix should be reduced before flattening. Based on our pressure test for the working test computer, which uses the CICIDS 2018 training dataset as shown in Table \ref{tab:cicids_2017}, our current computer capability can handle about $800$ host features. The goal is to condense the original message matrix of $100 \times 768$ into a smaller matrix $C^{'}$ with a $p \times q$ dimension. Furthermore, $C^{'}$ should have the number of message features $p \times q$ equal to $800$. To determine the most appropriate dimension, we evaluated various combinations of integers $p$ and $q$ that satisfy $p \times q$ not being greater than $800$. We used XGBoost to evaluate the detection performance with various matrices ($C^{'}$) of message features, keeping the same flow features ($132$) and event features ($226$). The message matrix was reduced to $10*80$, $15*50$, and $20*40$. The detection performance under different message matrices is shown in Fig. \ref{fig:DT_diff_row_col}. It was observed that the performance was almost the same under different message matrices. Therefore, we selected the message feature matrix $15*50$ as a representative in the subsequent tests.

\begin{figure*}[!hbt]
    \centering
    \includegraphics[width = 1\textwidth, trim=0cm 0cm 0cm 0cm,clip]{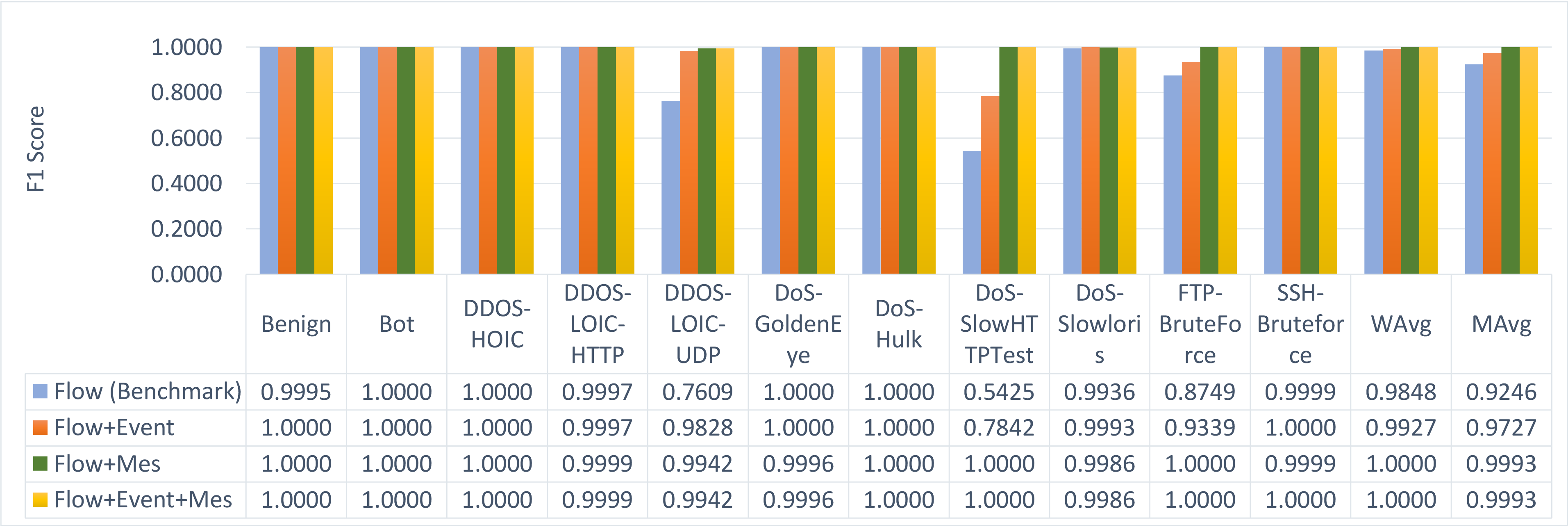}
    \caption{NIDS (flow-based) and hybrid NIDS and HIDS (flow, event, and message-based) performance comparison under CICIDS 2018. 'Mes' abbreviates message. Flow-based NIDS system performance is the benchmark against which the proposed hybrid flow and host features approaches are compared.}
    \label{fig:flow_host_res_com}
\end{figure*}

Fig. \ref{fig:flow_host_res_com} shows XGBoost (i.e., F1 score) results for four scenarios: 1) only flow feature $A$, 2) a combination of flow and event features $H_{1}$, 3) a combination of flow and message features $H_{2}$, and 4) a combination of flow, event, and message features $H_{3}$. We used XGBoost with only flow features as the benchmark for comparison with our proposed hybrid flow and host-based performance.
The results show that integrating message and event features improves the detection performance of all classes that cannot be fully detected. Particularly for the DDOS-LOIC-UDP attack, F1 score increases from $0.7609$ (flow features only) to $0.9828$ (with event features) and $0.9942$ (by integrating message and event features). The most significant improvement is achieved for the detection of DoS-SlowHTTPTest attack, which increases from $0.5425$ to $1.0000$. Moreover, the overall performance is significantly increased. For instance, the macro F1 score is increased up to $0.9993$ with the event and message features, compared to $0.9246$ with flow features only. The host features, whether event or message, clearly improve the attack detection accuracy. Moreover, message features show further advantages in boosting the intrusion detection performance compared to the event features. When flow features and message features are combined, a macro F1 score of $0.9993$ is achieved, while integrating flow features and event features leads to a macro F1 score of $0.9727$. 

\subsection{Performance evaluation under NDSec-1 dataset}
\label{sec:results_NDSec}
We followed a similar rule to combine flow and host features for the NDSec-1 dataset, starting by reducing the event and message feature matrices into smaller ones and flattening the two-dimensional event and message feature matrices into vectors. As introduced in Section \ref{sec:ndsec_dt_introduction}, event features are stored in a $2182 \times 8$ matrix $B$, while message features are stored in a $512 \times 768$ matrix $C$. We used the random selection method to determine the smaller matrices for the NDSec-1 dataset, based on previous experience and methodology for the CICIDS 2018 evaluation. We selected the event feature matrix as $B^{'}$ and the message feature matrix as $500 \times 8$, and the message feature matrix as $C^{'}$ with dimensions $100 \times 768$ for the NDSec-1 dataset. We skip the detailed procedure in matrix size determination, which is almost the same as the CICIDS 2018 dataset, as illustrated in Section \ref{sec:cicidsresults}.
Meanwhile, numerical results using the CICIDS 2018 dataset indicate that the dimensions of the message feature matrix marginally impact the intrusion detection performance, as shown in Fig. \ref{fig:DT_diff_row_col}. Most of the event features are excluded if a small matrix, such as $10 \times 10$, is chosen instead of the original $100 \times 768$ event feature matrix. One might expect this to reduce the effectiveness of the detection performance. However, with a smaller matrix of event and message features, the number of host features is also significant, up to $80800$ with $t_{event}=4000$ and $t_{mes}=76800$. Therefore, we employed Principle Component Analysis (PCA) \cite{abdi2010principal} to reduce dimensions further for the mix of flow, event, and message features. 

\begin{figure}[!hbt]
    \centering
    \includegraphics[width = 0.7\textwidth, trim=0.5cm 0cm 1cm 0cm,clip]{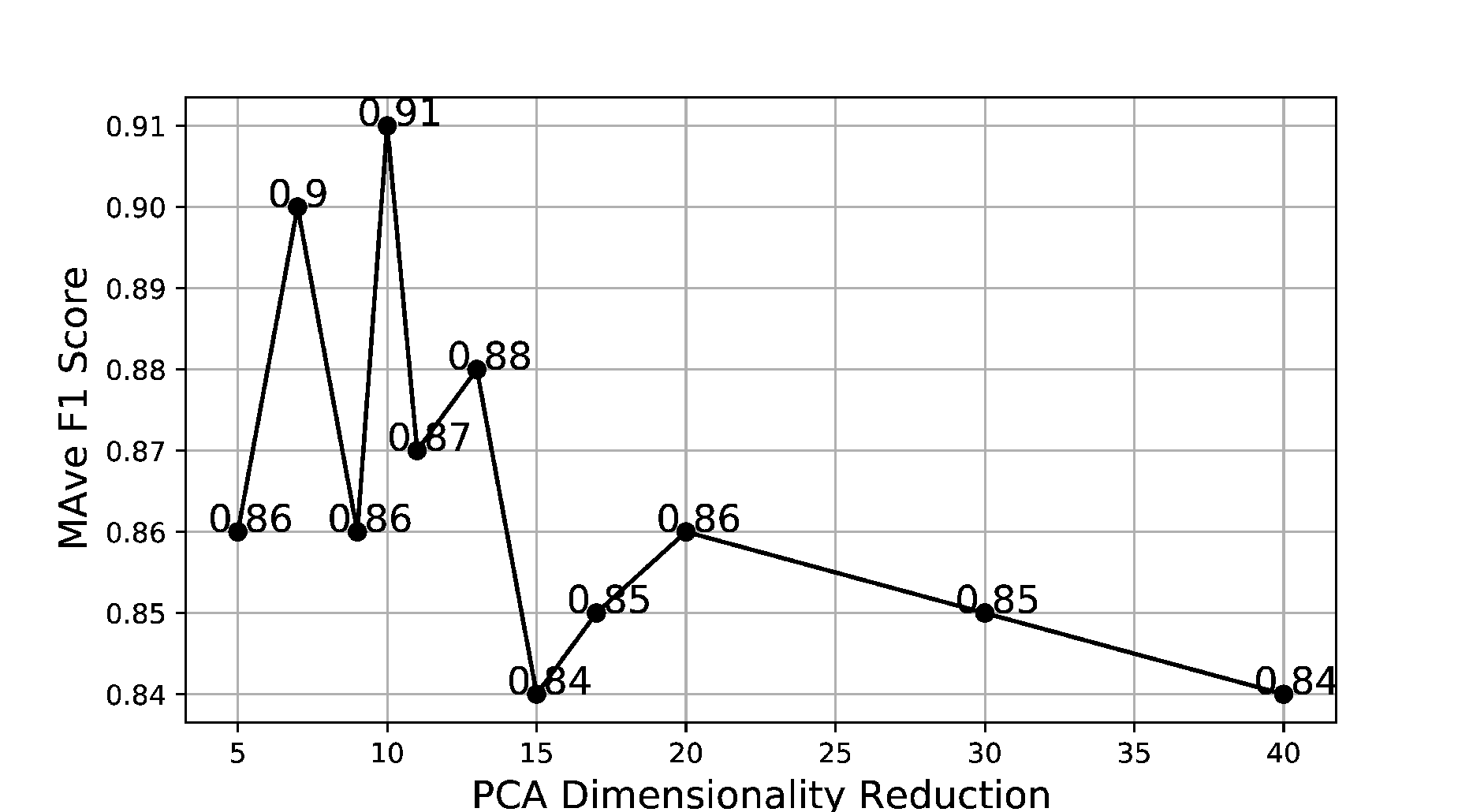}
    \caption{Comparison of the impact of various dimensions (reduced by PCA) under the NDSec-1}
    \label{fig:pca_comp}
\end{figure}

\begin{figure*}[!hbt]
    \centering
    \includegraphics[width = 1\textwidth, trim=0cm 0cm 0cm 0cm,clip]{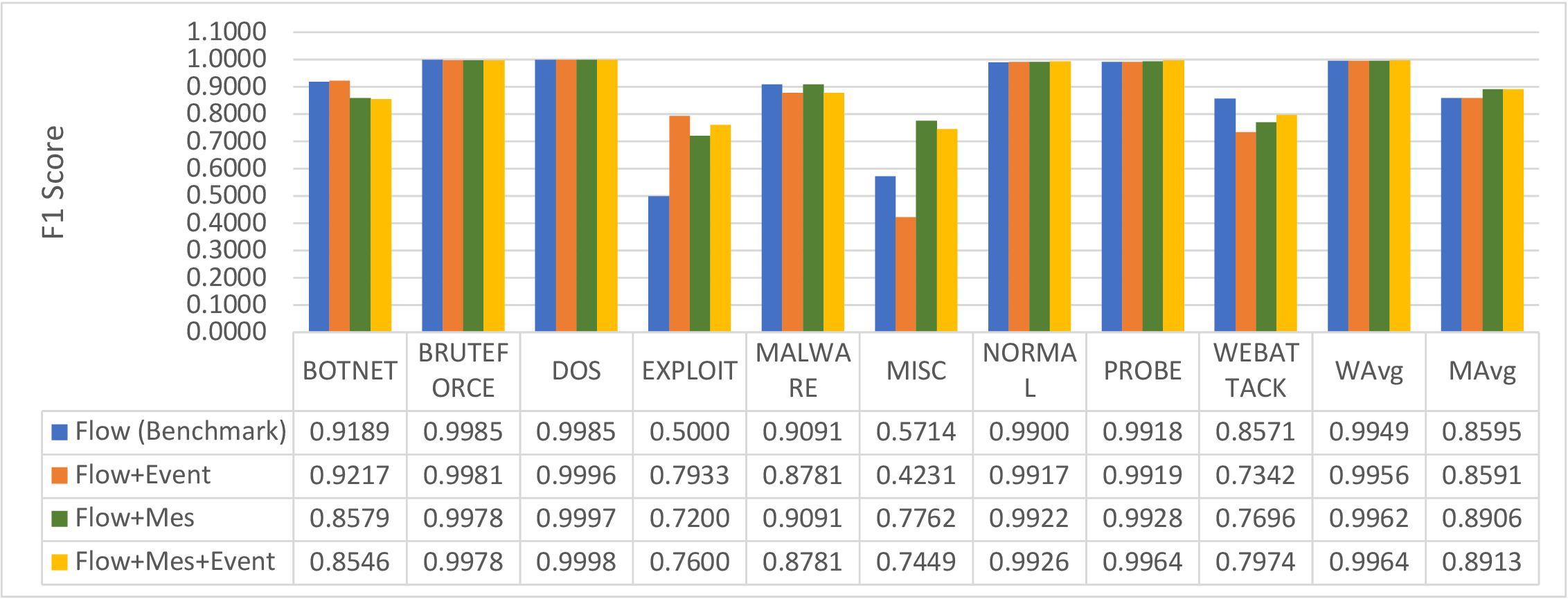}
    \caption{Performance comparison of NIDS (flow-based) and hybrid NIDS and HIDS (flow, event and message-based) under NDSec-1. 'Mes' abbrevaiates message. Flow-based NIDS system performance is the benchmark against which the proposed hybrid flow and host features approaches are compared.}
    \label{fig:newdt_res_comp}
\end{figure*}

\begin{figure}[!hbt]
    \centering
    \includegraphics[width = 0.8\textwidth, trim=2cm 7cm 2cm 2cm,clip]{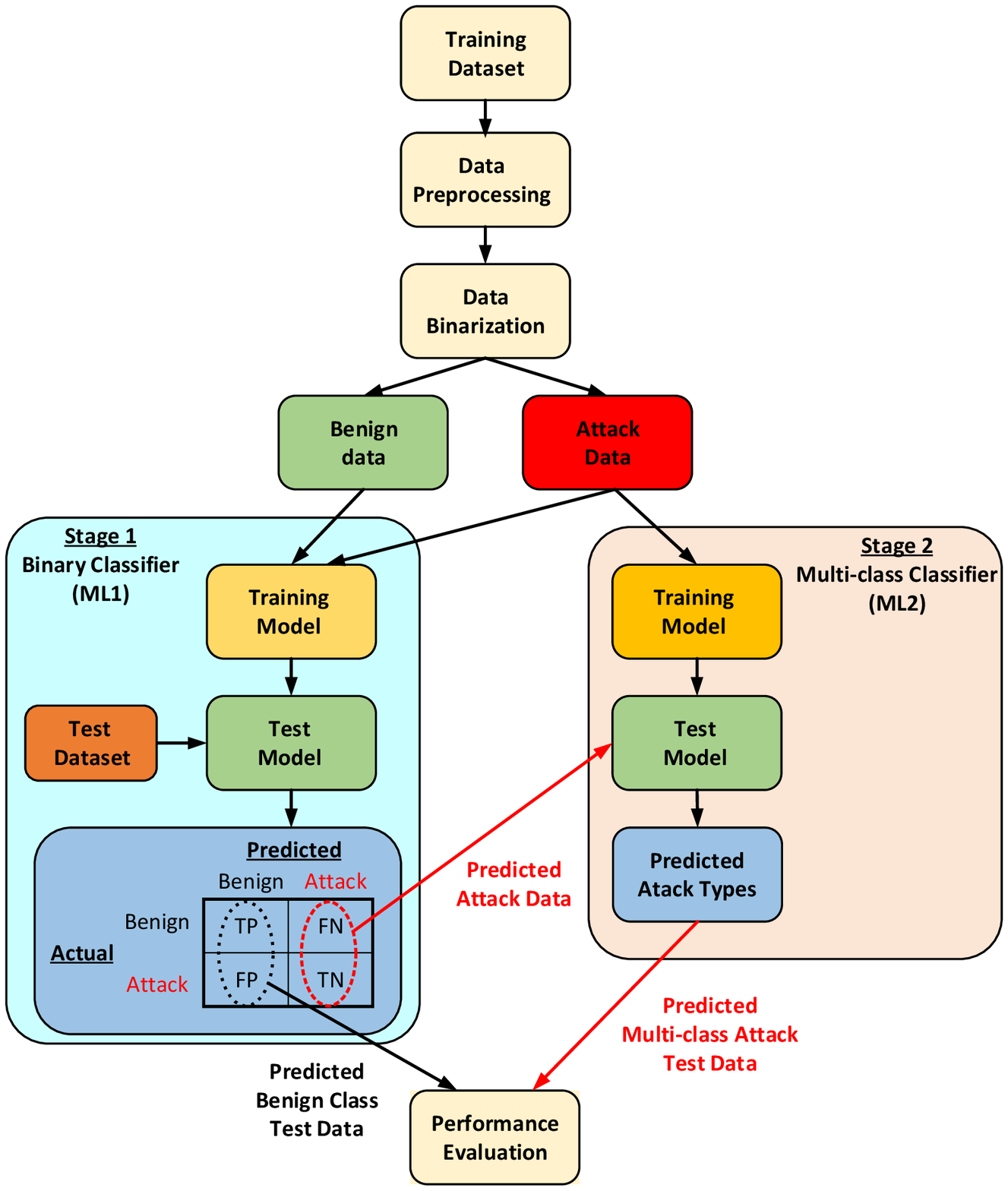}
    \caption{Two-stage collaborative classifier architecture. First stage builds on a binary classifier (ML1),  and the second stage builds on a multi-class classifier (ML2).}
    \label{fig:cascade_model_arch}
\end{figure}

Fig. \ref{fig:pca_comp} presents a comparison of performance results (F1 score) using PCA to reduce features to different dimensions. The highest F1 score is achieved when PCA reduces the features to $10$, with a macro F1 score of $0.91$. Therefore, in the following tests, PCA is applied to reduce the features to $10$.
Fig. \ref{fig:newdt_res_comp} presents the attack detection performance (F1 score) with flow and host features. The detection system using the composition of flow features, event features, and message features $H_{3}$ achieves the highest overall performance in macro average F1 score (MAvg) with $0.8913$ and a weighted average F1 score (WAvg) of $0.9964$. With $H_{3}$, the detection system demonstrates the best performance for DoS, Normal, and Probe. Specifically, MAvg is at $0.8913$ when compared to the flow-only case ($0.8595$), combined flow and event features $H_{1}$ ($0.8591$), and the combination of flow and message features $H_{2}$ ($0.8906$). When flow and event feature $H_{1}$ is utilized, the best performance for Botnet and Exploit is obtained with F1 scores of $0.9217$ and $0.7933$, respectively. Moreover, the flow and message combination $H_{2}$ case performs the best under Malware ($0.9091$) and Misc ($0.7762$). Thus, host features are helpful in boosting attack detection performance in terms of overall performance and most individual classes when compared to the case where only network features are used. However, introducing host features leads to a reduction in performance in the detection of Bruteforce and Webattack. For instance, Bruteforce F1 score reduces slightly from $0.9985$ (flow only) to $0.9981$ (flow and event), and Webattack reduces from $0.8571$ (flow only) to $0.7974$ (flow, event, and message together).

\section{Two-stage collaborative classifier}
\label{sec:cas}
\subsection{Method introduction}
\label{sec:cas_method}
The CICIDS 2018 dataset contains a majority of benign samples, with $308,375$ in the training dataset and $152,172$ in the test dataset, contributing to approximately $50\%$ in both datasets, as shown in Table \ref{tab:cicids_2017}. Based on the previous test results in Section \ref{sec:flat_method}, it can be observed that benign samples can be detected with an accuracy level close to $100\%$. Therefore, the detection performance of benign samples is almost unaffected by attack points.
Eliminating the benign samples would reduce the computational complexity of the intrusion detection system. Moreover, filtering out half of the samples in the dataset would reduce the dataset's complexity and improve the intrusion detection performance per attack type. Hence, we propose a two-stage collaborative classifier, as shown in Fig. \ref{fig:cascade_model_arch}, that integrates a binary classifier $ML1$ and a multi-class classifier $ML2$.

The proposed method conducts the training and testing procedures sequentially. In the first stage, the training dataset is used to train $ML1$. In the data preprocessing step, the original training dataset is labeled with two classes, including $0$ and $1$, representing benign and attack, respectively. All attack classes share one label as attack, while benign samples keep the same label as the original dataset in Table \ref{tab:cicids_2017}. $ML1$ is trained to discriminate between attack (label 1) and benign (label 0) classes.
In the second stage for $ML2$, the original training dataset filters out benign samples, and the filtered dataset is used to train $ML2$. Thus, $ML2$ is trained to discriminate between multiple attack types without benign samples, forming a multi-classifier. $ML2$ reduces the complexity of the problem and saves computing time.
The performance of the proposed two-stage collaborative classifier is verified by applying the test dataset. $ML1$ predicts whether a sample is benign or an attack. All samples estimated as intrusion samples are sent to $ML2$ for the second-stage classification. Performance evaluation of the proposed method is not straightforward since not all samples are predicted by one ML algorithm.
In this work, the overall performance combines $ML1$ and $ML2$ prediction results. Specifically, we follow $ML1$ estimation for benign samples and $ML2$ for attack sample prediction. This approach enables the detection of different types of intrusions while maintaining high accuracy in identifying benign traffic.

\begin{figure}[!hbt]
    \centering
    \includegraphics[width = 0.8\textwidth, trim=2cm 5cm 1.2cm 2.5cm,clip]{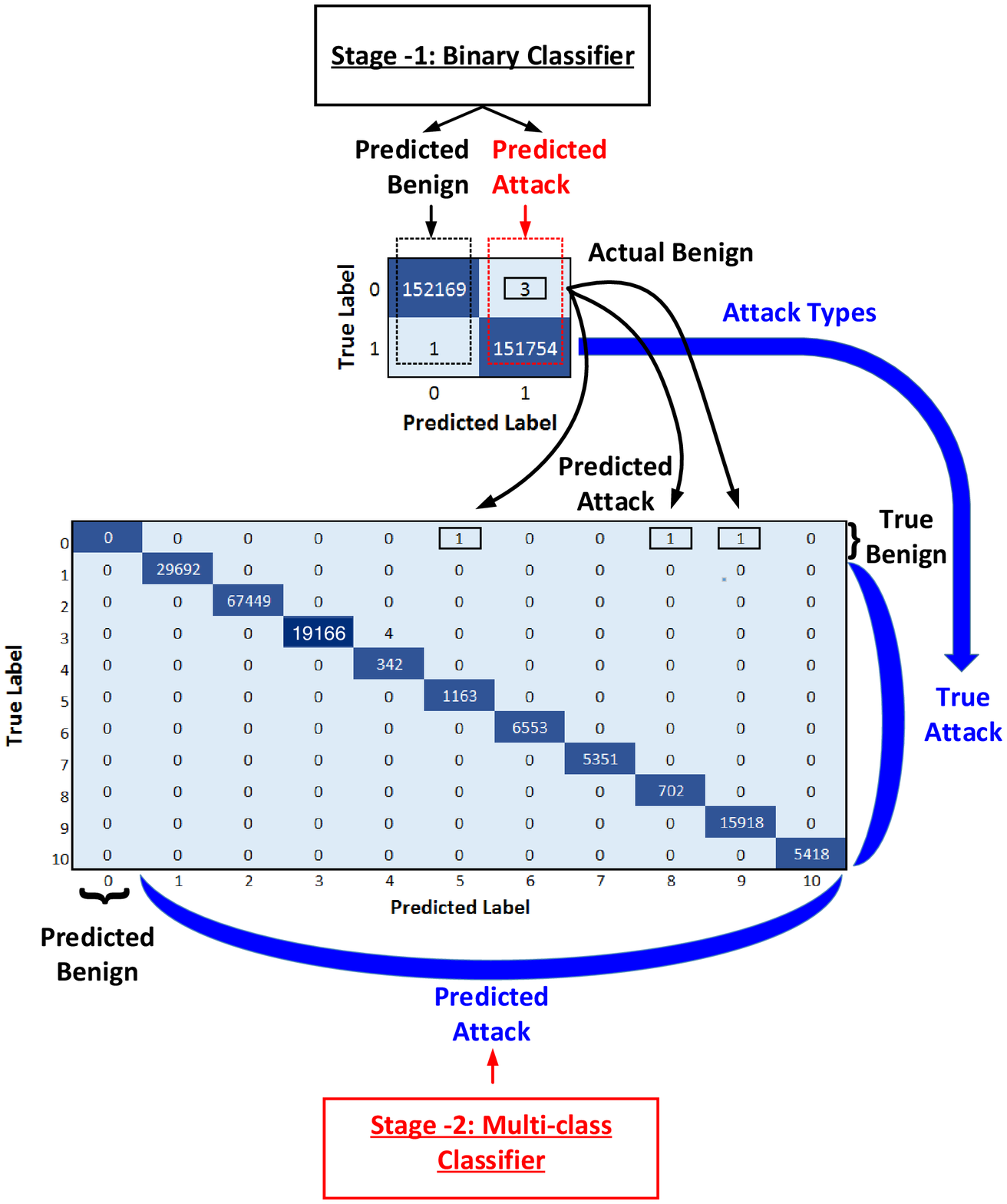}
    \caption{The confusion matrix of stage-1 (Binary Classifier) and   stage-2 (Mult-class Classifier) to demonstrate the prediction performance of the proposed collaborative two-stage method.}
    \label{fig:ml1_confusion}
\end{figure}

\begin{figure*}[!hbt]
\centering
    \begin{subfigure}{0.9\textwidth}
        \centering
        \includegraphics[width = 1\textwidth, trim=0cm 6cm 0cm 5cm,clip]{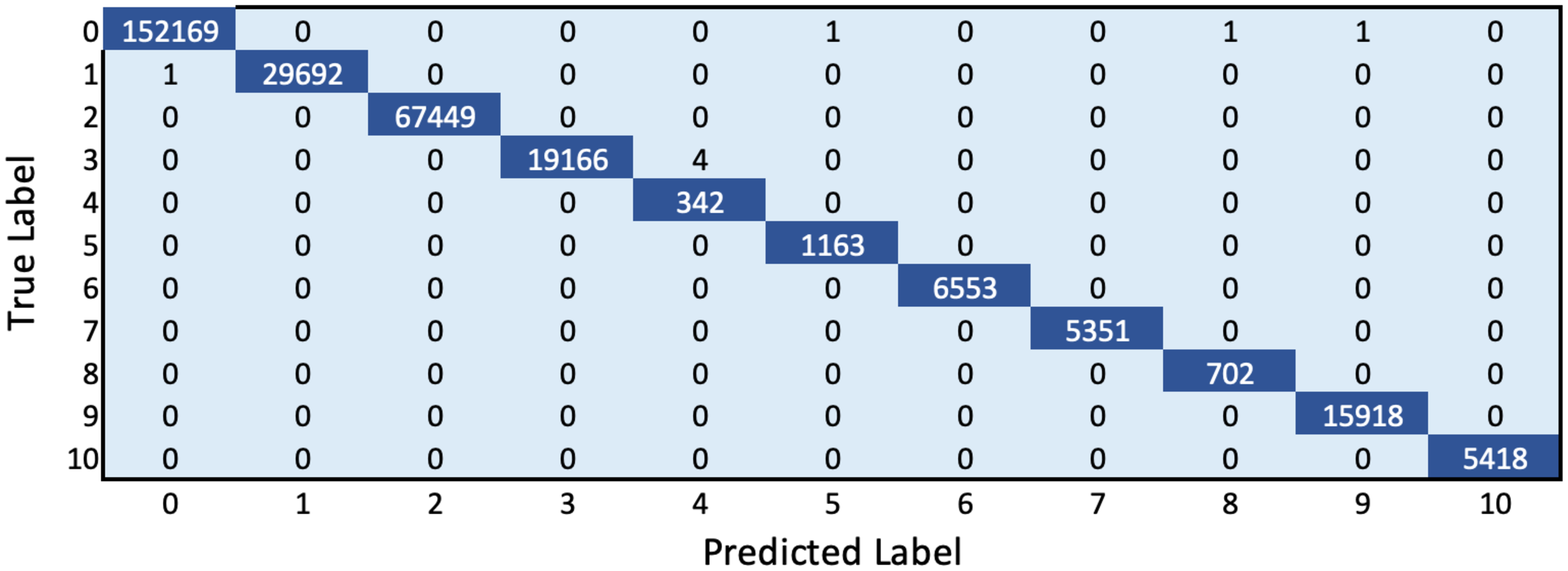}
    \caption{Proposed model with flow, event and message features}
    \label{fig:cascade_conf}
    \end{subfigure}
    \begin{subfigure}{0.9\textwidth}
        \centering
       \includegraphics[width = 1\textwidth, trim=0cm 5cm 0cm 5cm,clip]{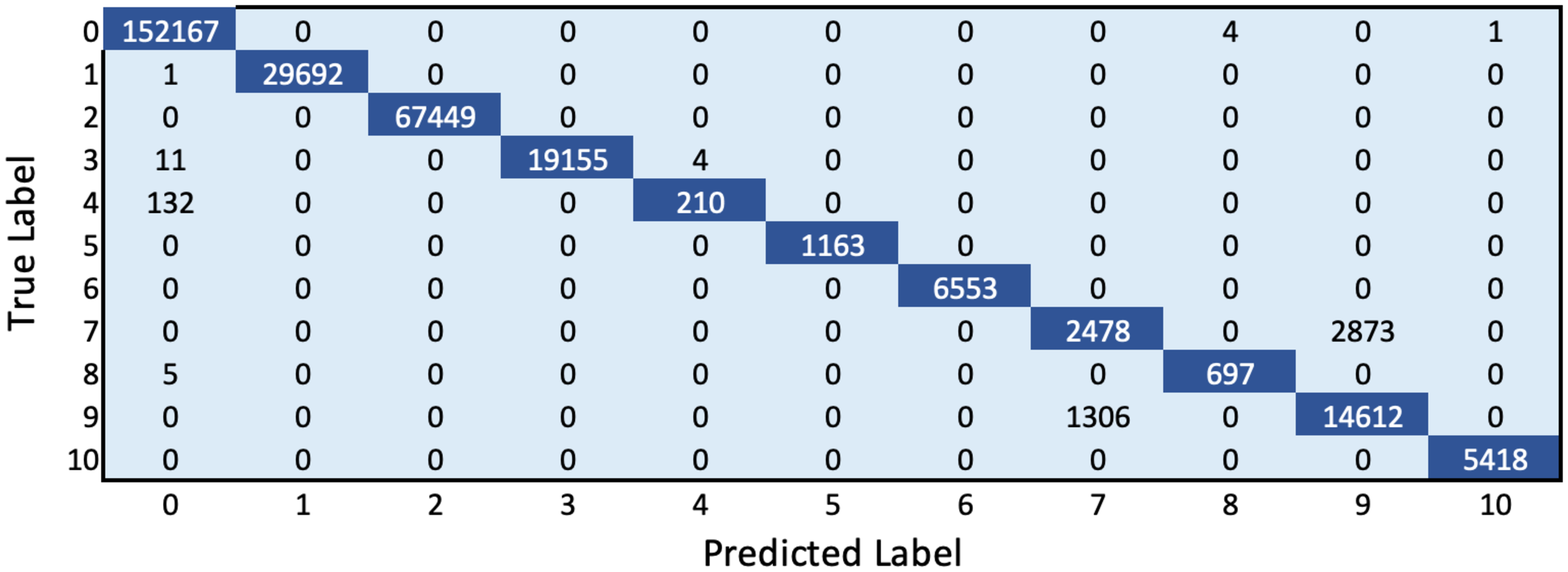}
    \caption{XGBoost with flow features}
    \label{fig:confusion_xgb_cascade}
    \end{subfigure}
    \caption{Confusion matrix for XGBoost using flow feature and the proposed two-stage collaborative classifier with flow, event and message features}
	\label{fig:binconfusionmatrixcomp}
\end{figure*}

\subsection{Performance evaluation}
The proposed two-stage classifier is evaluated using the CICIDS 2018 dataset. To compare with the previous work in Section \ref{sec:flat_method}, all $132$ flow features, flattened event features ($224$), and the same matrix of message features ($15*50$) are used to evaluate the proposed approach. Based on previous test results, XGBoost achieves almost $100\%$ accuracy for benign samples, making it the chosen base classifier in the two-stage collaborative classifier, as shown in Fig. \ref{fig:cascade_model_arch}. However, it is worth noting that the base classifier can be replaced with any other machine learning model. The objective of this study is not to propose a new machine learning algorithm but a new framework that can use any machine learning algorithm to boost the detection accuracy of multiple intrusive patterns. The presented method relies on an ML algorithm that achieves extremely high detection accuracy for benign samples. It uses a first-layer classifier, $ML1$, to identify and eliminate benign samples, and a second-layer classifier, $ML2$, to detect different types of intrusions. Using deep learning-based approaches is a prospective research direction to find a suitable $ML1$ instead of XGBoost.

\begin{figure}[!hbt]
    \centering
    \includegraphics[width = 0.8\textwidth, trim=0cm 0cm 0cm 0cm,clip]{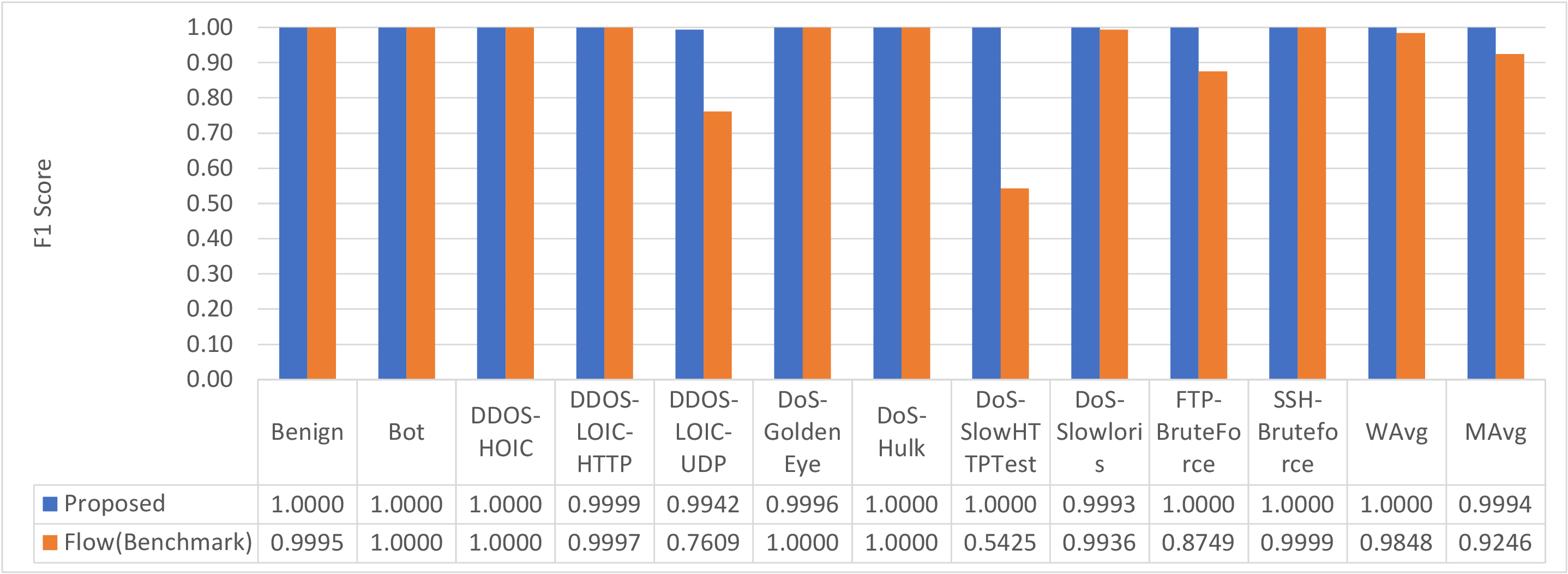}
    \caption{Performance comparison for the two-stage collaborative classifier under the CICIDS 2018 dataset The benchmark is the same one as shown in Fig. \ref{fig:flow_host_res_com} , using flow-only features, while the proposed method uses flow, event, and host features.}
    \label{fig:bin_mul_comp_xgb}
\end{figure}

When the test dataset is re-labeled for attack samples using one label instead of the original labels, the performance of XGBoost ($ML1$) can be seen in the confusion matrix illustrated in Fig. \ref{fig:ml1_confusion}. The confusion matrix shows that most benign samples (152169/152172) are predicted correctly and filtered by XGBoost (ML1). Only $3$ benign samples are predicted as attacks and sent to $ML2$ for classification. However, one (1) attack sample is predicted as benign, which is removed by $ML1$ prior to $ML2$. Fig. \ref{fig:ml1_confusion} demonstrates the classification results of the second-stage machine learning model ($ML2$) using XGBoost. Fig. \ref{fig:binconfusionmatrixcomp} demonstrates confusion matrix for the two methods. We can see the proposed method improves attack detection rate especially for DoS-LOIC-UDP (class 4), DoS-SlowHTTPTest (class 7) and FTPBruteForce (class 9). Fig. \ref{fig:bin_mul_comp_xgb} shows the performance comparison between the proposed two-stage collaborative classifier and XGBoost for individual classes and overall performance (macro average F1 score) by integrating $ML1$ and $ML2$ prediction results. 
The proposed approach significantly improves the macro average F1 score from $0.9246$ to $0.9994$, resulting in an $8.1\%$ improvement. It also enhances the detection performance for most single classes, especially for DoS-LOIC-UDP and DoS-SlowHTTPTest, with $30.7\%$ and $84.3\%$ improvements, respectively. However, the introduced  approach slightly reduces the performance of DDoS-GoldenEye, with a $0.04\%$ reduction from $1$ to $0.9996$. This is an unintended side effect of the proposed method.
Furthermore, the two-stage collaborative classifier only slightly improves performance compared to traditional ML presented in Section \ref{sec:cicidsresults}, from $0.9993$ to $0.9994$ in terms of macro average F1 score. As the performance is close to $100\%$, it is challenging to achieve significant improvements. Nonetheless, the proposed framework has the potential to improve the performance of intrusion detection systems by reducing computational complexity and improving the detection accuracy of multiple intrusive patterns.

\section{Conclusion and future work}
\label{sec:conclusion}
Network-based Intrusion Detection Systems (NIDS) have limitations in recognizing particular attack types, such as Advanced Persistent Threats, which require bridging with Host-based Network Intrusion Detection systems (HIDS). In this article, we proposed combining HIDS and NIDS to detect various types of intrusive patterns in a networked environment using Machine Learning (ML)-based approaches. We combined network-based and host-based features using a feature flattening approach and studied the impact of dimension reduction on message features at the hosts. We used traditional ML algorithms (e.g., XGBoost) and a two-stage collaborative classifierto detect intrusions. The first estimation stage used a binary classifier to discriminate benign and attack traffic, and the second stage used a multi-class classifier to discriminate multiple attack types. We evaluated our approach using two public datasets, CICIDS 2018 and NDSec-1, which contain network information and host-based data (e.g., event data and message data in host resources).
Our numerical results showed that the hybrid of network and host features significantly improved attack detection performance. The overall performance (macro average F1 score) increased from $0.9246$ to $0.9993$ under CICIDS 2018, representing an $8.1\%$ enhancement, and the detection performance for all individual attack types improved. For instance, the F1 score of DDOS-LOIC-UDP increased from $0.7609$ to $0.9942$, and DoS-SlowHTTPTest improved from $0.5425$ to approximately $1$ under the CICIDS 2018 dataset. Under the NDSec-1 dataset, the hybrid of flow and host features improved the overall detection performance (macro average F1 score) from $0.8595$ to $0.8913$. The two-stage collaborative classifier dramatically boosted performance, with the macro average F1 score increasing from $0.9246$ to $0.9994$ and $30.7\%$ and $84.3\%$ improvements in individual class performance for DoS-LOIC-UDP and DoS-SlowHTTPTest, respectively.
Our ongoing work includes evaluating graph-based datasets (e.g., StreamSpot, Long-hour dataset) introduced in \cite{han2020unicorn} using the presented method in this work. We also plan to deploy deep learning network-based approaches in the two-stage collaborative classifier instead of XGBoost to further boost detection accuracy.

\begin{acks}
This work is supported in part by the Ontario Centre for Innovation (OCI) under ENCQOR 5G Project $\#$31993.
\end{acks}

\bibliographystyle{ACM-Reference-Format}


\end{document}